\documentclass[10pt,journal,compsoc]{IEEEtran}
\ifCLASSOPTIONcompsoc
  \usepackage[nocompress]{cite}
\else
  \usepackage{cite}
\fi
\ifCLASSINFOpdf
\usepackage[pdftex]{graphicx}
\else
\usepackage[dvips]{graphicx}
\fi
\usepackage{array}
\usepackage{url}
\usepackage{color, colortbl}

\usepackage{enumitem}% http://ctan.org/pkg/enumitem

\usepackage{multirow}
\usepackage{booktabs}

\usepackage{hyperref}

\begin{document}
\title{Task-Based Effectiveness of Interactive Contiguous Area Cartograms}

\author{Ian~K.~Duncan, Shi~Tingsheng, Simon~T.~Perrault, and
  Michael~T.~Gastner
  \IEEEcompsocitemizethanks{
    \IEEEcompsocthanksitem{
      I.~K.~Duncan, S.~Tingsheng, and M.~T.~Gastner are with Yale-NUS
      College, Singapore 138527.
    }
    \IEEEcompsocthanksitem{
      S.~T.~Perrault is with the Singapore University of Technology
      and Design, Singapore 487372.
    }
  }
}


\markboth{IEEE Transactions on Visualization and Computer Graphics}%
{Duncan \MakeLowercase{\textit{et al.}}: Task-Based Effectiveness of Interactive Contiguous Cartograms}



\IEEEtitleabstractindextext{%
\begin{abstract}

Cartograms are map-based data visualizations in which the area of each map region is proportional to an associated numeric data value (e.g., population or gross domestic product).
A cartogram is called contiguous if it conforms to this area principle while also keeping neighboring regions connected.
Because of their distorted appearance, contiguous cartograms have been criticized as difficult to read.
Some authors have suggested that cartograms may be more legible if they are accompanied by interactive features (e.g., animations, linked brushing, or infotips).
We conducted an experiment to evaluate this claim.
Participants had to perform visual analysis tasks with interactive and noninteractive contiguous cartograms.
The task types covered various aspects of cartogram readability, ranging from elementary lookup tasks to synoptic tasks (i.e., tasks in which participants had to summarize high-level differences between two cartograms). 
Elementary tasks were carried out equally well with and without interactivity.
Synoptic tasks, by contrast, were more difficult without interactive features.
With access to interactivity, however, most participants answered even synoptic questions correctly.
In a subsequent survey, participants rated the interactive features as ``easy to use'' and ``helpful.''
Our study suggests that interactivity has the potential to make contiguous cartograms accessible even for those readers who are unfamiliar with interactive computer graphics or do not have a prior affinity to working with maps.
Among the interactive features, animations had the strongest positive effect, so we recommend them as a minimum of interactivity when contiguous cartograms are displayed on a computer screen.
\end{abstract}

}

\maketitle

% is not selected <OR> if conference mode is selected - because compsoc
% conference papers position the abstract like regular (non-compsoc)
\IEEEdisplaynontitleabstractindextext

% \fi
\IEEEpeerreviewmaketitle

\ifCLASSOPTIONcompsoc
\IEEEraisesectionheading{\section{Introduction}\label{sec:introduction}}
\else
\section{Introduction}
\label{sec:introduction}
\fi

\begin{figure*}[!t]
\centering
\includegraphics[width=7in]{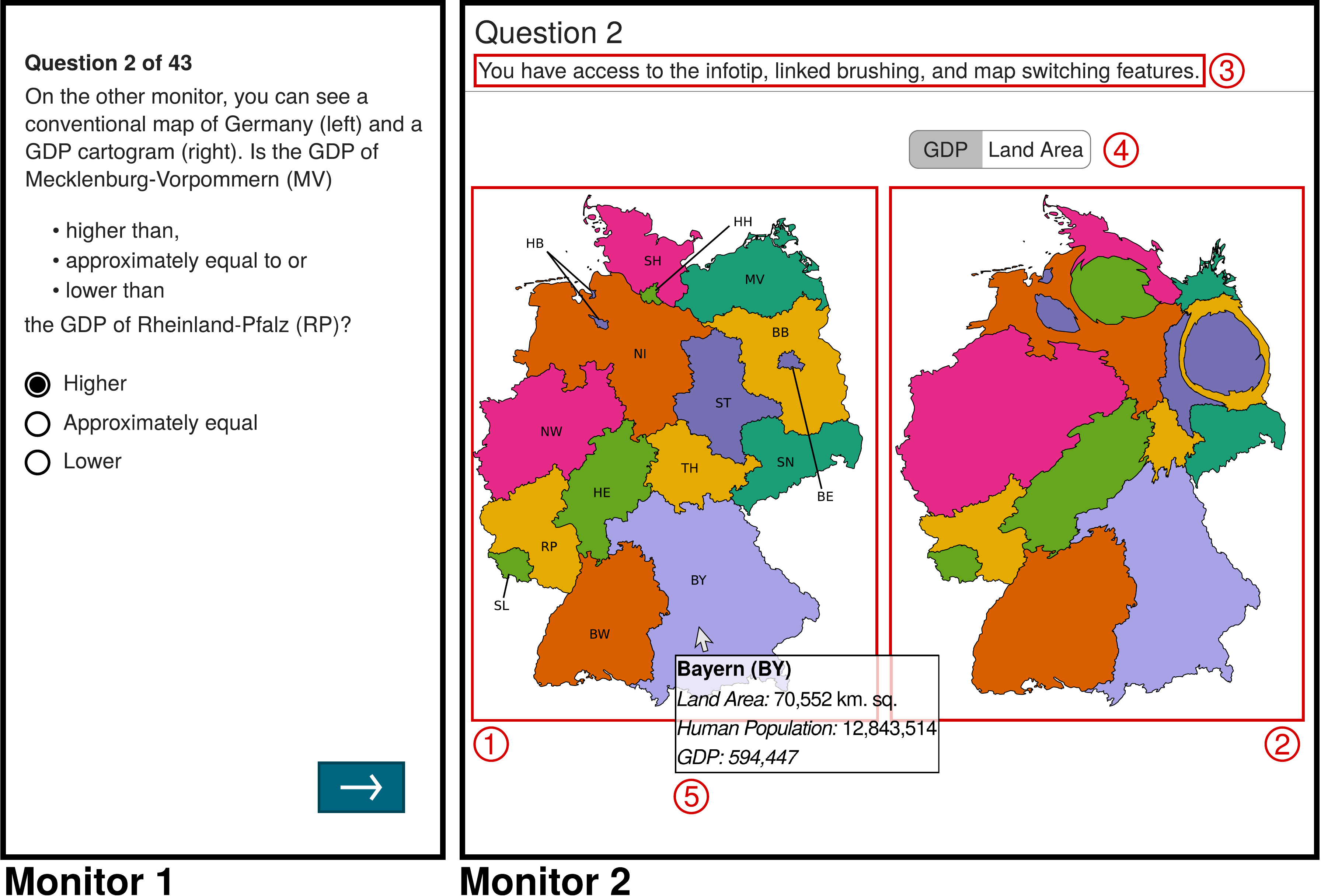}
\caption{Screenshot of the two-monitor setup used by participants to complete the map reading tasks.
We rescaled the aspect ratios of the screens to fit the page width.
Here we use a \emph{Compare} task for Germany as an example (see Section~\ref{subsec:tasks} for the task description). On Monitor 1, participants read the current task and entered their answer. On Monitor 2, participants were presented with the cartogram viewing interface containing (1) a labeled conventional map and (2) a contiguous cartogram of the same country.
Corresponding regions appear in the same color on the conventional map and the cartogram.
Participants were informed of (3) which interactive features they could use to complete the current task.
For some tasks, they could also switch the map displayed on the right to a cartogram based on another statistic using (4) the
map switching selector.
One of the buttons that could be selected was titled ``Land Area.''
When this button was pressed, the map displayed on the right switched to an equal-area map, which can be considered as a cartogram where the statistic displayed is the land area of each administrative region.
Hovering the mouse over a map region triggered (5)
an infotip containing the region's name followed by the numeric data used to generate
the displayed cartograms.} % Removed extra white-space from fig.
\label{fig:cartview}
\end{figure*}

\IEEEPARstart{T}{he} amount of geospatial information stored in digital databases and shared over the Internet is growing rapidly.
To communicate information contained in geospatial data to a wide audience, we need effective visualization tools.
Cartograms have emerged as an alternative to traditional thematic mapping techniques, such as choropleth maps, proportional symbol maps, and dot-density maps~\cite{dent_cartography_2008, kronenfeld_visualizing_2017}.
The Worldmapper~\cite{gotthard_worldmapper_nodate, hennig_remapping_2010} and Londonmapper projects~\cite{dorling_london_2014}, for example, make extensive use of cartograms.
In a cartogram, each region is depicted by an area that is proportional to its corresponding value in the statistical data set used.
In Fig.~\ref{fig:cartview}, we illustrate this idea using data for the gross domestic products (GDPs) of federal states in Germany.
The left map in Fig.~\ref{fig:cartview} is a conventional equal-area map, where Berlin (labeled as BE) occupies only 4.8\% of the area of Saxony (SN).
In the cartogram (the map on the right of Fig.~\ref{fig:cartview}), however, Berlin's area appears 12\% larger than Saxony's so that the proportions of the states' GDPs can be correctly represented.

In this article, we concentrate on cartograms that are contiguous.
That is, if map regions share a common border in a geographic space (i.e., on a conventional map) then they are also neighbors on the cartogram, and vice versa.
The cartogram in Fig.~\ref{fig:cartview}, for example, is contiguous.
Many other cartogram types exist (e.g.,
mosaic cartograms~\cite{cano_mosaic_2015}, Olson's noncontiguous~\cite{olson_noncontiguous_1976}, and Dorling's circular
cartograms~\cite{dorling_area_1996}).
Previous surveys and experiments have compared the effectiveness of
different cartogram types~\cite{sun_effectiveness_2010,
  nusrat_evaluating_2018}.
Contiguous cartograms were consistently among the participants'
preferred visualizations across a variety of contexts and tasks, even
if other cartogram types might have been more suitable for specific
applications or user groups.

Cartograms first gained popularity in the early 20th
century~\cite{raisz_rectangular_1934} when they were hand-drawn
and intended for inclusion in print media.
Like most other forms of data visualization, currently cartograms are
more frequently generated electronically and viewed on a computer
screen than printed on paper.
With the increasing presence of cartograms on the World Wide Web, news media have also
adopted them to support their online content.
Various newspapers have posted cartograms of election
forecasts~\cite{the_new_york_times_electoral_2012, zitner_draw_2016},
election results~\cite{los_angeles_times_u.s._2012,
  the_telegraph_election_2015, franklin_eu_2016}, and other demographic statistics~\cite{fairfield_changing_2008, byron_map_2008} on their
websites.
Despite the popularity of cartograms, they inevitably appear distorted
when compared to conventional maps.
For this reason, some authors have questioned whether cartograms are
legible and comprehensible for the average
reader~\cite{rittschof_learning_1996, fotheringham_quantitative_2007,
  eckert_site_2008, roth_value-by-alpha_2010}.
If a cartogram is displayed on a computer screen, it is possible to include interactive features to alleviate these concerns.
Indeed, many of the cartograms posted by the news media include features such as a button or slider to switch between different cartograms and a conventional map~\cite{zitner_draw_2016, fairfield_changing_2008, byron_map_2008, clark_carbon_nodate, bbc_news_results_2019}, or a mouse-over effect to highlight regions~\cite{franklin_eu_2016,kommenda_election_2019}, or infotips~\cite{byron_map_2008, the_new_york_times_electoral_2012, los_angeles_times_u.s._2012, zitner_draw_2016, franklin_eu_2016, the_telegraph_election_2015, clark_carbon_nodate, bbc_news_results_2019, kommenda_election_2019}.
However, it remains an open research question whether interactivity makes it easier to understand the information shown in a cartogram~\cite{nunez_hungarian_2015}.

The purpose of this study is to evaluate whether cartograms can be used to communicate information more effectively if the viewer can interact with them using the following three interactive features proposed in the previous cartogram literature~\cite{ware_using_1998}, \cite{tobler_thirty_2004}, \cite{herzog_developing_2005}:

\begin{itemize}
\item \textbf{Cartogram-switching animation}: The user can choose between different data sets by clicking a button on the screen (Fig.~\ref{fig:cartview}). As the user selects a new data set, the previously displayed cartogram morphs into a new cartogram. In our implementation, the transition from one cartogram to another is achieved by smoothly moving the polygon vertices from the start to end position in a one-second time interval.

\item \textbf{Linked brushing:} The cartogram is displayed alongside a conventional map, as shown in Fig.~\ref{fig:cartview}. As the participant hovers the mouse over a region on the cartogram, the corresponding region is simultaneously highlighted on the conventional map and vice versa.
In our experiment, we highlighted the region by increasing the brightness of its fill color, but other forms of highlighting (e.g., changing the border color or thickness) are also conceivable.

\item \textbf{Infotip:} As the participant hovers the mouse over a map region, a pop-up appears at the location of the cursor (see Fig.~\ref{fig:cartview}).
The text in the pop-up contains the region's name and the data (e.g., GDP) represented by the corresponding area in the cartogram.
\end{itemize}

We judge the effect of these interactive features on map-reading tasks, which range from elementary lookup tasks to high-level synoptic summaries of the cartograms, drawn from an objective-based task taxonomy~\cite{nusrat_task_2015}.
Results from our experiment indicate that interactive features can help readers perform certain data analysis tasks more accurately than on cartograms without such features.
While the inclusion of interactive features may not have an observable effect on simple tasks (with $<20\%$ error rates), we did notice significant improvements for synoptic tasks, where cartogram-switching animations dramatically improved the accuracy of the participants' responses.
Cartogram-switching animations may thus effectively remove concerns about the legibility and effectiveness of contiguous cartograms in the previous literature~\cite{eckert_site_2008, woodruff_i_2008}.

At the end of the experiment, participants filled out an attitude
survey.
Overall, they expressed strongly positive opinions about all the interactive features tested in the experiment.
In addition to the longstanding recommendations of
presenting cartograms with a legend and alongside a conventional map~\cite{dent_communication_1975}, we therefore suggest that cartograms be presented with interactive features to improve reader comprehension.

\section{Related Work}
\subsection{Previous Evaluations of Cartograms}

The utility of cartograms has been debated for several decades.
On one hand, Dent already reported in 1972 that students found cartograms intriguing and were eager to experiment with them~\cite{dent_note_1972}.
On the other hand, he also criticized cartograms as bordering on the ``surrealistic''~\cite{dent_communication_1975} and primarily being used for their shock value~\cite{dent_cartography_2008}.
In 1975, he conducted a series of experiments, which were among the earliest attempts to objectively evaluate cartogram effectiveness~\cite{dent_communication_1975}.
Participants were asked to estimate the population in the northeastern United States either from a cartogram or a proportional symbol map based on a conventional map projection.
The accuracy of the participants' estimates was comparable for both map types.
At the end of the test, participants characterized the cartogram as ``innovative'' and ``interesting,'' but also pointed out that they found it difficult to read.

In a similar experiment in 1983, Griffin measured the accuracy and speed with which participants identified regions on a cartogram shown alongside a conventional map of Adelaide's electoral subdivisions~\cite{griffin_recognition_1983}.
One region was highlighted on one of the maps, and the task was to find the corresponding region on the other map. When the distortions were greater, the task became more challenging. Griffin hypothesized that participants struggled to establish the mental transformation between conventional maps and their cartograms.
In a similar experiment, Kaspar \textit{et al.}~\cite{kaspar_empirical_2011} noticed that the difficulty of the spatial inference task also depended on the shape of the polygon on a conventional map.
Polygons with regular shapes (e.g., rectangles) appeared subjectively more distorted on the cartogram than polygons with irregular shapes.

At least for simple tasks, Kaspar \textit{et al.} still concluded that cartograms could be as effective and efficient as traditional graduated circle maps.
Most other cognitive experiments have echoed this result~\cite{aschwanden_kognitionsstudien_1998, sun_effectiveness_2010, han_experimental_2017}, whereas Gao \textit{et al.}~\cite{gao_usability_2018} argue that value-by-alpha maps~\cite{roth_value-by-alpha_2010} or proportional symbol maps are at least as effective, efficient, and popular with users as cartograms.
Comparisons between different cartogram methods have shown that university students can cope well with the distortions inherent in contiguous cartograms~\cite{nusrat_evaluating_2018}.
Secondary school students, however, preferred the simpler geometries depicted in rectangular or mosaic cartograms~\cite{nunez_hungarian_2015, markowska_using_2019}.

All the experiments mentioned in this section were done with static cartograms.
Because many cartograms are currently shown on websites rather than in print, one may wonder whether interactivity (e.g., implemented with D3.js~\cite{bostock_d_2011}) fundamentally alters the viewers' attitudes towards cartograms.
As Goodchild noted about map displays in general, ``it is unreasonable that a technology optimized under the narrow constraints of pen and paper would turn out to be indistinguishable from one optimized under the much broader constraints of digital technology''~\cite{goodchild_stepping_1988}.

\subsection{Interactivity in General Mapping Software}
Interactive graphics have been included in general-purpose statistical software (e.g., for exploratory data analysis~\cite{cleveland_dynamic_1988}) and geographic software (e.g., GIS and web-based mapping services~\cite{peterson_interactive_1995}) for a long time. 
Zooming, for example, is a quintessential interactive feature in mapping software that encourages the viewer to focus on small portions of a map~\cite{crampton_interactivity_2002, persson_towards_2006}. 
According to Harrower and Sheesley~\cite{harrower_designing_2005}, cartographers should consider incorporating zooming into interactive mapping because the feature allows for greater information density and is commonly understood by map readers. 
Apart from zooming, there are interactive features permitting comparisons between two or more maps that represent the same region, such as translucent overlays, a blending lens, and swiping~\cite{lobo_evaluation_2015}. Previous studies have exhibited examples of how these features are implemented on a computer display~\cite{gough_method_nodate}.
Several experiments concluded that interactivity has a positive impact on performance for map-related tasks~\cite{oviatt_multimodal_1997, brock_interactivity_2015, knight_interactivity_2017, alencar_de_mendonca_does_2019}, although the type of interactivity and the type of tasks involved vary greatly between these experiments.
Experiments by Keehner \textit{et al.}~\cite{keehner_spatial08} underline that the value of interactive features in spatial cognition tasks is not the interactivity in itself, but the easier access to informative views of the spatial objects.

It is surprising that the literature on cartograms, which are at the intersection of statistical and geographic visualization~\cite{raisz_general_1938}, has so far largely ignored the opportunities of interactivity. 
One reason for this neglect might be that the most discussed interactive features in geographic visualization are not directly useful when applied to cartograms. 
For example, while zooming is helpful in displaying the boundaries of the map in greater detail, the main purpose of a cartogram---as well as most other infographic
maps~\cite{smith_online_2016}---is to show the relative importance of regions in a larger geographic context with a fixed spatial extent.
For instance, zooming into the cartogram on the right of Fig.~\ref{fig:cartview} would not help us relate Berlin's GDP to the total GDP of Germany.
Other features are better suited for map comparison than zooming, but they often rely on the assumption that the two maps that are to be compared are based on the same underlying map projection.
For contiguous cartograms, this assumption is invalid because different input statistics produce different cartogram projections.
Therefore, many tools for map comparisons, such as translucent overlays, a blending lens, or swiping,
would leave the viewer confused about the relation between the incongruent boundaries on the two cartogram layers.
The only potentially promising map comparison technique that we found in the literature was map morphing, which does not require the same projection for both maps.
Map morphing continuously interpolates between the projections of two incongruent maps with a short animation~\cite{li_animating_2008, pantazis_are_2009, lobo_animation_2019}.
Reilly and Inkpen~\cite{reilly_map_2004} reported that map morphing significantly improved accuracy for tasks that included the comparisons of region sizes on different projections.
They noted that morphing also felt effective and enjoyable to the participants because it was easy to follow shifts in positions during the animation.

\subsection{Interactivity in Cartograms}

In addition to reviewing the general literature that covers the broad topic of interactivity in maps, we also thoroughly reviewed the more specific literature about cartograms, examined the available cartogram software, and inspected many cartograms posted on the World Wide Web.
All interactive features that were mentioned or implemented fall into three categories: animations, linked brushing, and infotips.
%Other interactive features mentioned in the general literature on data visualization or cartography (e.g., zooming, panning) would not add obvious value to cartograms.
Therefore, we restricted our experiment to these three features for which we found a precedent in previous cartogram research.

The idea of combining animations with cartograms had been discussed since the
mid-1980s~\cite{monmonier_technological_1985, campbell_animated_1990, dibiase_animation_1992}, but it remained a theoretical prospect until Dorling produced an animated time series of UK election cartograms in 1992~\cite{dorling_stretching_1992}.
Each constituency was represented by an arrow on a circular cartogram.
The color of the arrow illustrated the vote composition, and the direction indicated the vote swing.
The animation was played from a videotape so that viewers could not directly interact with the graphics.
The first experiment in which participants could interact with a cartogram animation on a computer screen was conducted by Ware in 1998~\cite{ware_using_1998}.
The animation showed a morphing transition from a conventional map to a contiguous cartogram.
Participants were assigned to different groups.
Some groups could play, pause, or rewind the animation, whereas the control group could only view still images.
All participants had to identify regions highlighted on a conventional map by clicking on the corresponding region on a cartogram.
Participants with access to the animation needed more time to complete the tasks, but were more likely to give correct answers, especially when they were unfamiliar with the geography of the displayed country.
Ware hypothesized that these participants spent some of their time rewinding the animation to ensure that their response was correct.
Overall, she strongly advocated the use of animated presentations of cartograms, pointing out that user satisfaction is more important than a quick response time.

Instead of gradually deforming one map into another, it is also possible to compare maps by juxtaposition.
Placing two maps alongside each other can be particularly effective in combination with linked brushing, whereby pointing at a position on one of the maps simultaneously highlights the corresponding position on the other map~\cite{lobo_evaluation_2015}.
Inspired by implementations of scatterplot brushing in statistical software~\cite{newton_graphics_1978, becker_brushing_1987} and its generalization to geographic data~\cite{monmonier_geographic_1989, andrienko_interactive_1999}, linked brushing has been proposed by several authors as a method of highlighting the correspondence between a conventional map and its cartogram~\cite{dykes_exploring_1997, haining_spatial_2003, tobler_thirty_2004}. Dykes commented that linked brushing ``reduces the oft-quoted difficulties in relating cartogram symbols with the places that they represent.''~\cite{dykes_cartographic_1998}.
Linked cartogram brushing has been implemented by the GeoViz Toolkit~\cite{hardisty_geoviz_2011} and Tableau~\cite{mundigl_cartograms_2015}.
However, as Tobler noted, the efficacy of linked cartogram brushing has so far not been evaluated~\cite{tobler_thirty_2004}.

Besides visually highlighting the region under the mouse pointer, Nusrat \textit{et al.}~\cite{nusrat_state_2016} hypothesized that cartograms may also benefit from an infotip that reveals the exact value of the numeric data represented by the region's area.
Infotips have been implemented in cartogram software such as MAPresso~\cite{herzog_developing_2005} and are included in many cartograms posted on the World Wide Web~\cite{byron_map_2008, the_new_york_times_electoral_2012, los_angeles_times_u.s._2012, office_for_national_statistics_visualising_2015, spruijt_bouncymaps_2017}.
In the context of general map applications, experiments have found that users preferred infotips when tasked to retrieve background information about objects on a map even if other possibilities were available~\cite{heidmann_interactive_2003}.
However, Brath and Banissi have remarked that an infotip may increase user response times without providing compensatory benefits in understanding and readability~\cite{brath_bertins_2018}.
When unnecessary for completing a given data analysis task, an infotip may act as an intrusive visual stimulus that makes it more difficult for users to keep their attention on relevant aspects of a data visualization, potentially increasing the response time and decreasing the accuracy~\cite{proctor_human_2003}. Despite these concerns, we are not aware of any previous experimental assessments of infotips in the context of cartograms.

\subsection{Evaluating Interactivity with Task Taxonomies}

While a general theory of interactive cartography is still in its infancy, there is broad agreement that we should assess the quality of a cartographic visualization with well-defined task taxonomies~\cite{wehrend_appendix_1993, zhou_visual_1998, amar_low-level_2005}.
Roth distinguishes between objective-based taxonomies, which define tasks in terms of verbs that imply user intent (e.g., ``identify,'' ``compare''), and operator or operand-based taxonomies, which categorize tasks by the specific features or visualizations used~\cite{roth_cartographic_2012}.
From a card-sorting study with expert interactive map users, Roth identified five general objective primitives for interactive geovisualization: ``identify,'' ``compare,'' ``rank,'' ``associate,'' and ``delineate''~\cite{roth_empirically-derived_2013}.
However, he admits that there was a large amount of variation in the participants' opinions, so it is unclear whether these objectives are optimal for user studies of all forms of geovisualization.

Here, we adopt the objective-based taxonomy that Nusrat and Kobourov developed specifically for cartograms~\cite{nusrat_task_2015}.
Their suite of tasks is not explicitly designed to evaluate interactivity but can be easily modified to test the interactive features in our experiment.
By adopting the objective-based perspective, we treat the interactive features as conditions under which the objectives can be reached, but participants are free to choose whether they use the implemented features during the experiment.

\section{Experiment}

\subsection{Tasks}
\label{subsec:tasks}

From the cartogram tasks proposed by Nusrat and Kobourov~\cite{nusrat_task_2015},
we chose eight general categories that are relevant in the context of interactive cartograms. We list the task types together with example tasks in Table~\ref{tab:tasktypes}. In the typology of~\cite{andrienko_exploratory_2006}, the first seven tasks are ``elementary.''
That is, they are basic statistical and map reading tasks that refer to individual subregions (e.g., states or provinces).
By contrast, the last task in Table~\ref{tab:tasktypes} (``Summarize'') is ``synoptic'': it is a complex task that required participants to analyze and compare the whole set of regions on multiple cartograms that were jointly displayed on the same screen.

\definecolor{LightGrey}{rgb}{0.9,0.9,0.9}
\definecolor{MidGrey}{rgb}{0.8,0.8,0.8}

\begin{table*}[!tp]
\renewcommand{\arraystretch}{1.2}
\normalsize
\caption{Eight task types used in our experiment with example tasks for each type}
\label{tab:tasktypes}
\centering
% 1. Cluster
\begin{center}
\begin{tabular*}{\textwidth}[h]{ | p{4em} | p{47.7em}|}
\hline
\rowcolor{LightGrey}
\multicolumn{2}{|c|}{Cluster}\\
\hline
Task & For a given region, participants were required to select the region with the most similar area on the displayed cartogram from a set of four possible candidate regions. Here and in all other tasks, the task description was shown on monitor 1. The cartogram was displayed on monitor 2 (see Fig.~\ref{fig:cartview}).\\
\hline
Example & On monitor 2, you can see a conventional map of Brazil (left) and a cattle population cartogram (right). Out of the states listed below, which has a cattle population most similar to Mato Grosso do Sul (MS)? \\[.4em]
\hline
% 2. Compare
\rowcolor{LightGrey}
\multicolumn{2}{|c|}{Compare}\\
\hline
Task & For two given regions, $R_1$ and $R_2$, participants were required to decide if the area of $R_1$ was higher than, lower than, or approximately equal to the area of $R_2$ on the displayed cartogram.
 \\
\hline
Example & On monitor 2, you can see a conventional map of Germany (left) and a GDP cartogram (right). Is the GDP of Mecklenburg-Vorpommern (MV) higher, lower, or approximately equal to the GDP of Rheinland-Pfalz (RP)? \\[.4em]
\hline
% 3. Detect Change
\rowcolor{LightGrey}
\multicolumn{2}{|c|}{Detect Change}\\
\hline
Task & For a given region and two cartograms $C_1$ and $C_2$, participants were required to decide if its area in $C_1$ was higher than, lower than, or approximately equal to its area in $C_2$.\\
\hline
Example & On monitor 2, you can see a conventional map of Brazil (left), a cattle population cartogram, and a human population cartogram (right). Is the area of Amazonas (AM) in the cattle population cartogram higher, lower, or approximately equal to its area in the human population cartogram?\\[.4em]
\hline
% 4. Filter
\rowcolor{LightGrey}
\multicolumn{2}{|c|}{Filter}\\
\hline
Task & For a given region $R$, participants were required to select a map region with an area greater than that of $R$ on the displayed cartogram.\\
\hline
Example & On monitor 2, you can see a conventional map of Germany (left) and a population cartogram (right). Out of the states listed below, which one(s) have a population higher than Baden-W\"urttemberg (BW)? There may be more than one correct answer.\\[.4em]
\hline
% 5. Find Adjacency
\rowcolor{LightGrey}
\multicolumn{2}{|c|}{Find Adjacency}\\
\hline
Task & For a given region, which was highlighted on a cartogram displayed on monitor 1, participants were required to identify which regions are neighbors from a set of four candidates.
Participants had to infer the answer from maps displayed on monitor 2.\\
\hline
Example & On monitor 2, you can see a conventional map of Brazil (left) and a population cartogram (right). Which state(s) are neighbors of the state highlighted in red in the population cartogram below (i.e., on monitor 1)? There may be more than one correct answer.\\[.4em]
\hline
% 6. Find Top
\rowcolor{LightGrey}
\multicolumn{2}{|c|}{Find Top}\\
\hline
Task & Participants were required to identify the region with the largest area in the displayed cartogram. \\
\hline
Example & On monitor 2, you can see a conventional map of the USA (left) and a crop production cartogram (right). Which state has the highest crop production?\\[.4em]
\hline
% 7. Recognize (Recognize)
\rowcolor{LightGrey}
\multicolumn{2}{|c|}{Recognize}\\
\hline
Task & For a given region, which was highlighted on a cartogram displayed on monitor 1, participants were required to identify its name based on maps shown on monitor 2.\\
\hline
Example & On monitor 2, you can see a conventional map of India (left) and a population cartogram (right). What is the name of the state highlighted in red in the population cartogram below (i.e., on monitor 1)?\\[.4em]
\hline
% 8. Summarize
\rowcolor{LightGrey}
\multicolumn{2}{|c|}{Summarize}\\
\hline
Task & Regions were partitioned into three zones, which were colored yellow, purple, and pink.
For two given cartograms, $C_1$ and $C_2$, participants were required to identify whether the yellow, purple, and pink zones increased in area, decreased in area, or showed no noticeable change in area between $C_1$ and $C_2$. \\
\hline
Example & On monitor 2, you can see a conventional map of Germany (left) and population cartograms for the years 1985 and 2015 (right). Three different regions are highlighted in yellow, purple, and pink. What can you say about the trend in population growth between 1985 and 2015? \\
\hline
\end{tabular*}
\end{center}
\end{table*}

\subsection{Data Sets}
\label{subsec:datasets}

\begin{figure*}[!t]
\centering
\includegraphics[width=7in]{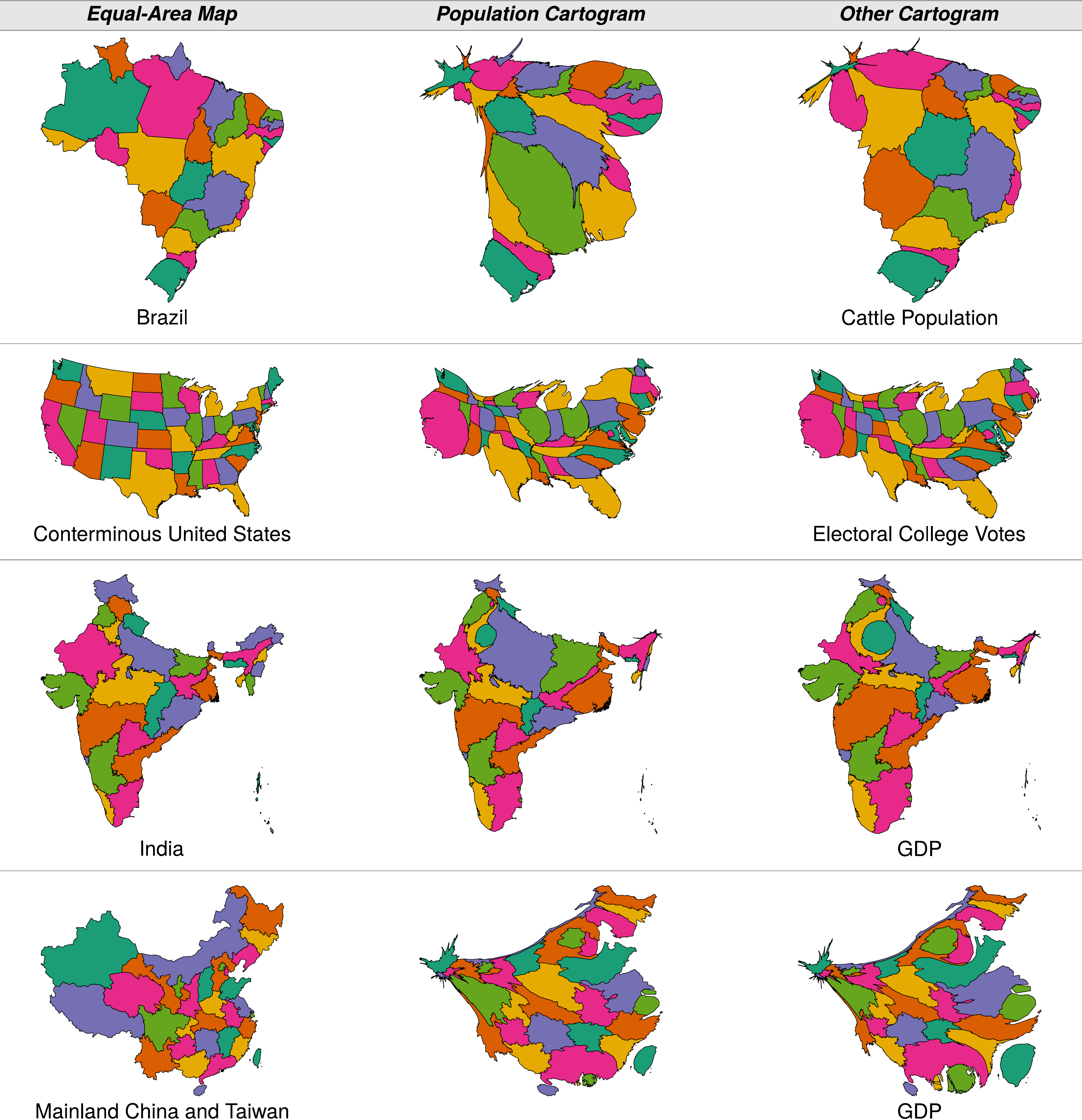}
\caption{Selection of cartograms shown during the experiment.} % Removed extra white-space from fig.
\label{fig:cartview_gallery}
\end{figure*}

For each task, we generated cartograms of five different parts of the world:
\begin{itemize}
    \item Germany (all 16 Bundesl\"ander),
    \item Brazil (all 26 states and the Federal District),
    \item the conterminous United States (48 states and Washington, D.C.),
    \item India (all 28 states and 7 union territories excluding Lakshadweep),
    \item mainland China, the Hong Kong and Macao Special Administrative Regions, and Taiwan (total of 34 administrative units).
\end{itemize}
Representative cartograms are shown in Fig.~\ref{fig:cartview} (Germany) and Fig.~\ref{fig:cartview_gallery} (Brazil, United States, India, and China).
We opted for real countries with recognizable outer boundaries rather than artificial geometries or deliberately unidentifiable subsets of census areas~\cite{beecham_map_2017} because we wanted the experimental tasks to resemble realistic use cases of cartograms.
We controlled for possible previous knowledge bias by treating the regions as a blocking factor in the experimental setup (see Section~\ref{sec:design}).

The cartogram areas represented statistics from a variety of data sets.
Apart from population and GDP data, we also visualized data that we expected few participants to be familiar with (e.g., agricultural production by state in the US, cattle production by state in Brazil).
For \textit{Summarize} tasks, we used actual or predicted population data from two different years.
All the cartograms were produced with the fast flow-based method~\cite{gastner_fast_2018}.
In each task, we presented a map made with a conventional equal-area projection alongside the cartogram (Fig.~\ref{fig:cartview}).

On the conventional map, we identified all regions with two-letter abbreviations---an example is shown in Fig.~\ref{fig:cartview}---to simplify the task for participants who were unfamiliar with the geography of the displayed country.
Unlike the conventional maps, none of the cartograms showed the regions' abbreviations because those labels would have defeated the purpose of some task types (e.g., \textit{Recognize}).
As a visual hint to the participants, the colors of matching regions on the conventional maps and their corresponding cartograms were identical.

\subsection{Participants}
We recruited 55 participants.
They were all students or staff of the National University of Singapore.
Seventeen of the participants were female; 38 were male.
The age range was from 18 to 52 years (mean 21.8, standard deviation 4.6).
The participants received 10 SGD (around 7.05 USD) as compensation for their time.
Because the experiment required the participants to distinguish between map regions highlighted with different colors, we used an Ishihara test to determine whether any participants were color blind. One participant exhibited potential color blindness.
The performance of this participant did not differ significantly from that of the other participants. Hence, we included this participant's responses in our analysis.

Because almost all participants were university students, we acknowledge that our results may only apply to a younger, more educated group of people. Similar limitations apply to previous cartogram evaluations in the literature (e.g., \cite{dent_communication_1975, rittschof_learning_1996, kaspar_empirical_2011, nusrat_evaluating_2018}). However, given the simplicity of the tasks, we believe that our results can be generalized to healthy adults and teenagers without any major vision impairment or reading disabilities.

\subsection{Procedure}

The participants were seated in front of two liquid-crystal display monitors, each with a resolution of 1920 $\times$ 1080.
On monitor 1, the participants read the task descriptions and entered their answers with the mouse or keyboard.
Monitor 2 displayed a graphical user interface that showed the conventional maps and cartograms.
The user interface was a web-based software application that allowed the user to trigger an animation by clicking a selector button (Fig.~\ref{fig:cartview}).
Linked brushing and infotips were enabled by hovering over a map region with the mouse.
We used Qualtrics XM to display the experiment tasks to the participants and collect their answers.

The experiment consisted of four parts.

\begin{enumerate}[label={(\arabic*)}]
\item
\label{experiment:intro}
\textbf{Introduction: }At the beginning of the experiment, all participants signed a form consenting to participate in the experiment. Then, they watched a five-minute video containing an introduction to cartograms and a description of the experiment. The participants were allowed to pause and rewind the video as they wished. They could also ask the experiment supervisor for clarification at any time. Afterwards, the participants had an opportunity to practice how to use cartogram-switching animations, linked brushing, and infotips.
For this practice run, we showed participants a conventional map of the conterminous United States alongside an interactive cartogram whose areas represented either electoral votes, population, or land area.
Participants did not have to complete any specific map reading task, but we required participants to confirm that they understood how each interactive feature worked before they were permitted to continue to the next stage of the experiment.
\item
\label{experiment:prelim}
\textbf{Preliminary questions: } We collected information about each participant's age, gender, and level of education. We also asked participants to judge their familiarity with maps, cartograms, and interactive computer graphics using a 5-point Likert scale.
Finally, we conducted an Ishihara color blindness test.
\item
\label{experiment:tasks}
\textbf{Cartogram tasks:} The participants answered 40 multiple choice questions that required them to analyze a conventional map and one to two cartograms.
If answer options consisted of named regions (i.e., for the task types \textit{Cluster}, \textit{Filter}, \textit{Find Adjacency}, \textit{Find Top}, and \textit{Recognize}), we selected distractors randomly, but we aimed to include at least one distractor that appeared plausible after a cursory glance.
Participants were informed that their responses and the amount of time taken to complete each task was recorded, but we also told them that there was no time limit for their answers.
Additionally, we recorded the computer screen so that we could analyze how participants used the interactive features provided.
\item
\label{experiment:attitude}
\textbf{Attitude study: }Adapting the semantic differential technique used by Dent~\cite{dent_communication_1975} and Nusrat \textit{et al.}~\cite{nusrat_evaluating_2018}, we asked participants to rate the aesthetics and effectiveness of the three interactive features evaluated in the experiment. We selected pairs of opposite words (e.g., ``conventional'' vs.\ ``innovative,'' ``hindering'' vs.\ ``helpful'').
For every pair of words, we asked participants to rate each interactive feature on a 5-point Likert scale.
Following the example of~\cite{nusrat_evaluating_2018}, participants did not have to give separate responses for each task type.
Hence, the participants responded three times to each word pair, once for each interactive feature. 
\end{enumerate}

The participants were supervised in person, in a one-on-one setting. All participants completed the experiment in around 50 minutes.
The instructional video used in part~\ref{experiment:intro} and the complete list of questions used in parts~\ref{experiment:prelim}--\ref{experiment:attitude} are available as supplemental material for this article.

\subsection{Design}
\label{sec:design}
We used an $8 \times 5$ within-subject experimental design with two independent variables: the eight task types listed in Table~\ref{tab:tasktypes} and five experimental conditions, which depended on the availability of interactive features during the task:
\begin{itemize}
\item
no interactivity,
\item
only a cartogram-switching animation is available,
\item
only linked brushing is available,
\item
only the infotip feature is available,
\item all three features (i.e., animation, linked brushing, and infotips) are available.
\end{itemize}
Every combination of a task type with one of these five conditions appeared exactly once during each session.
Thus, each participant completed exactly 40 trials during part~\ref{experiment:tasks} of the experiment.
Because each participant encountered each combination of a task type and a feature-condition only once, we cannot infer whether participants became more efficient at using the features during the experiment.
However, participants rated all features as ``easy to use'' at the end of the experiment (see Section~\ref{sec:user_preferences}), so we expect the learning curve to be almost flat.

\begin{figure}
\centering
\includegraphics[width=0.5\textwidth]{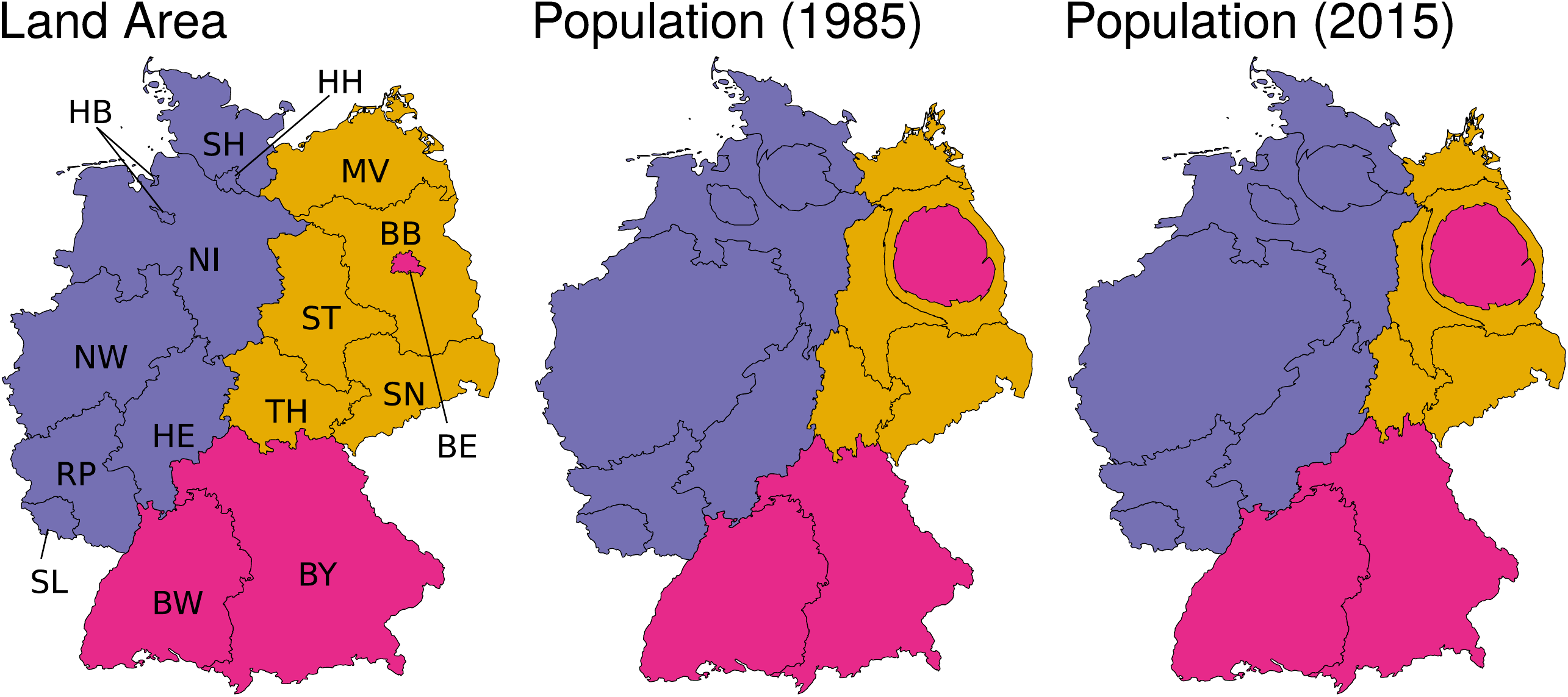}
\caption{Conventional map and cartograms presented to participants during the \textit{Summarize} task for Germany. The map was divided into three zones: purple, pink, and yellow. Participants were required to identify whether each zone increased, decreased, or did not change in area between the first (1985 population) and second (2015 population) cartogram.}
\label{fig:summarize_example}
\end{figure}

For most task types, the participants had to select one answer out of four choices.
However, for the \textit{Filter} and \textit{Find Adjacency} task types, participants could select multiple answers. For these questions, it was possible that more than one region matched the search criterion. All such regions had to be selected for the task to be completed correctly. \textit{Summarize} tasks were split into three sub-tasks (one for each colored zone, see Fig.~\ref{fig:summarize_example} for an example), but participants could only select one answer (``Growth,'' ``Approximately no change,'' ``Shrinking'') for each zone. For this task type, we considered a participant's trial as a success only if the answers to all three sub-tasks were correct. We also deemed ``Approximately no change'' as an alternative correct answer if the change in area was no larger than $1\%$. In a pilot study, we noticed that a difference in area of below $1\%$ is too subtle to be observed.
We discuss the effect of setting different thresholds in Section~\ref{subsec:performance_for_summarize}.

The order of the tasks was the same for all participants.
The order of the five interactive feature-conditions was counterbalanced using a Latin square.
% To prevent the order of interactive feature combinations within each task type from affecting the results, participants were assigned to one of five groups. Each group had the same number of participants.
% The order in which participants encountered the interactive features during the experiment was the same in each group but differed between the groups.
For the five tasks of the same type (e.g., the five tasks of the \textit{Cluster} type), we used each of the five parts of the world listed in Section~\ref{subsec:datasets} (i.e., a different region for each task) to reduce the potentially confounding effect of different levels of familiarity with the displayed maps.
All participants encountered the same random sequence of all 40 possible combinations of task types and parts of the world.
The five interactive-feature conditions appeared in a random permutation for the five tasks of the same type. In Section 1 of the online supplemental text, we include a table with the order in which participants encountered the combinations of map regions, task types, and interactive features.

\subsection{Data Analysis}
\label{sec:data_analysis}

For each task type in Table~\ref{tab:tasktypes}, we wanted to compare the error rates and response times under the five different conditions listed in Section~\ref{sec:design}.
To compare error rates, we treated the participants' responses as binary data (correct vs.\ incorrect) and used Cochran's Q test with the null hypothesis that the interactive features had no influence on the probability of giving a correct response.
The test statistic is $\chi^2$-distributed with $4$ degrees of freedom.
For post-hoc analysis, we used pairwise McNemar tests.
We applied the Bonferroni-Holm correction to adjust the $p$-values.
We consider the adjusted $p$-values as significant if they are below $0.05$.
We caution against overinterpreting $p$-values~\cite{cumming_understanding_2011} and, therefore, also state effect sizes and confidence intervals (CIs) for the post-hoc analysis.
For the McNemar test, we measured the effect size with the odds ratio and determined CIs with the method developed by Fay~\cite{fay_two-sided_2010}.
The null hypothesis corresponds to an odds ratio equal to $1$.

To compare the response times, we discarded all incorrect responses given by the participants.
Consequently, our time observations are not necessarily paired across the five interactivity conditions.
Even after discarding wrong responses, the distribution of response times is right-skewed.
Outliers with long response times were presumably a consequence of our instruction to the participant that they could take as long as they needed to answer each question.
With this instruction, we aimed to mimic a realistic scenario for reading cartograms in online news stories, where readers can look at cartograms without a fixed time limit.
Because of the outliers, the response time distributions for some tasks fail the Shapiro-Wilk test of normality.
Thus, we resorted to non-parametric Kruskal-Wallis tests to identify the main effects.
The test statistic of the Kruskal-Wallis tests is $\chi^2$-distributed with $4$ degrees of freedom.
For post-hoc analysis, we used pairwise Mann-Whitney U tests with the Bonferroni-Holm correction.
We express the effect size of the Mann-Whitney U tests in terms of the pseudomedian difference~\cite{hollander_nonparametric_2013}, which we denote by $\Delta$.

We have made the data and R scripts used for our statistical analysis publicly available at Zenodo~\cite{ian_k_duncan_2020_3832179}.

\subsection{Hypotheses}
\label{subsec:experiment_hypotheses}

Prior to the experiment, we expected interactive features to make some tasks in Table~\ref{tab:tasktypes} easier for the participants.
Our hypotheses were as follows.
%Since the main goal of a cartogram is to accurately convey information, we believe that the effects would be stronger on accuracy than time.

\subsubsection{Cartogram-switching Animations}
Cartogram-switching animations are useful to compare the evolution of different quantities over time~\cite{dorling_stretching_1992}.
For this reason, we believed that animations may greatly help participants to perform the \emph{Detect Change} and \emph{Summarize} tasks.
We foresaw a potentially positive effect on accuracy for these tasks, but the effect on the response time was unclear, as animations would also consume time.

\begin{itemize}
\item
\textbf{H1}: Participants will make fewer errors with cartogram-switching animations in \textit{Detect Change} and \textit{Summarize} tasks.
\end{itemize}

\subsubsection{Linked Brushing}
Linked brushing is useful for participants because it highlights a selected region simultaneously on a regular map and a corresponding cartogram, allowing participants to quickly locate regions of interest on both representations.
We believed that this feature could help participants to be more accurate when they execute \textit{Find Adjacency}, \textit{Find Top}, and \textit{Recognize} tasks, because they can first identify the relevant region on the cartogram on monitor 2 by comparing it with the region's shape or size on monitor 1, then hover the mouse over this region on monitor 2, and finally read off the correct answer from the labeled conventional map.
We also hypothesized that participants might be faster when using linked brushing.

\begin{itemize}
\item \textbf{H2-a}: Participants will make fewer errors with linked brushing in \textit{Find Adjacency}, \textit{Find Top}, and \textit{Recognize} tasks.
\item \textbf{H2-b}: Participants will need less time to perform these tasks using linked brushing.
\end{itemize}

\subsubsection{Infotips}
Infotips impart precise numeric information when the mouse hovers over a specific region, but reading the text in the infotip takes time.
Interacting with infotips also demands from the users that they carefully control how the cursor moves between different parts of the maps.
In addition, the infotip may occlude some parts of the maps, which could slow down map reading in general.
Therefore, we expected that infotips would increase execution time.
However, we believed that infotips could improve accuracy for all tasks.
The only exception is the synoptic task \textit{Summarize}, where information about small-scale individual regions is not directly relevant.

\begin{itemize}
\item
\textbf{H3-a}: Participants will make fewer errors with infotips in all elementary tasks compared to the no-interactivity condition.
\item
\textbf{H3-b}: Participants will need \emph{more} time when using infotips.
\end{itemize}

\subsubsection{All Interactive Features}
Using all interactive features should, theoretically, provide participants with more tools and information to perform the task.
The three features do not visually interfere with each other, so we predicted that the all-features condition would lead to higher accuracy than all other conditions.
However, when all three features are active, participants must process more information, so we expected the increased accuracy to come at the cost of increased execution time.

\begin{itemize}
\item
\textbf{H4-a}: Unlike the no-interactivity condition, participants will make fewer errors for every task type when all interactive features are available.
\item
\textbf{H4-b}: Participants will generally need more time in the all-features condition than in the no-interactivity condition.
\end{itemize}

\section{Results}

\begin{figure*}[!tp]
\centering
\includegraphics[width=0.9\textwidth]{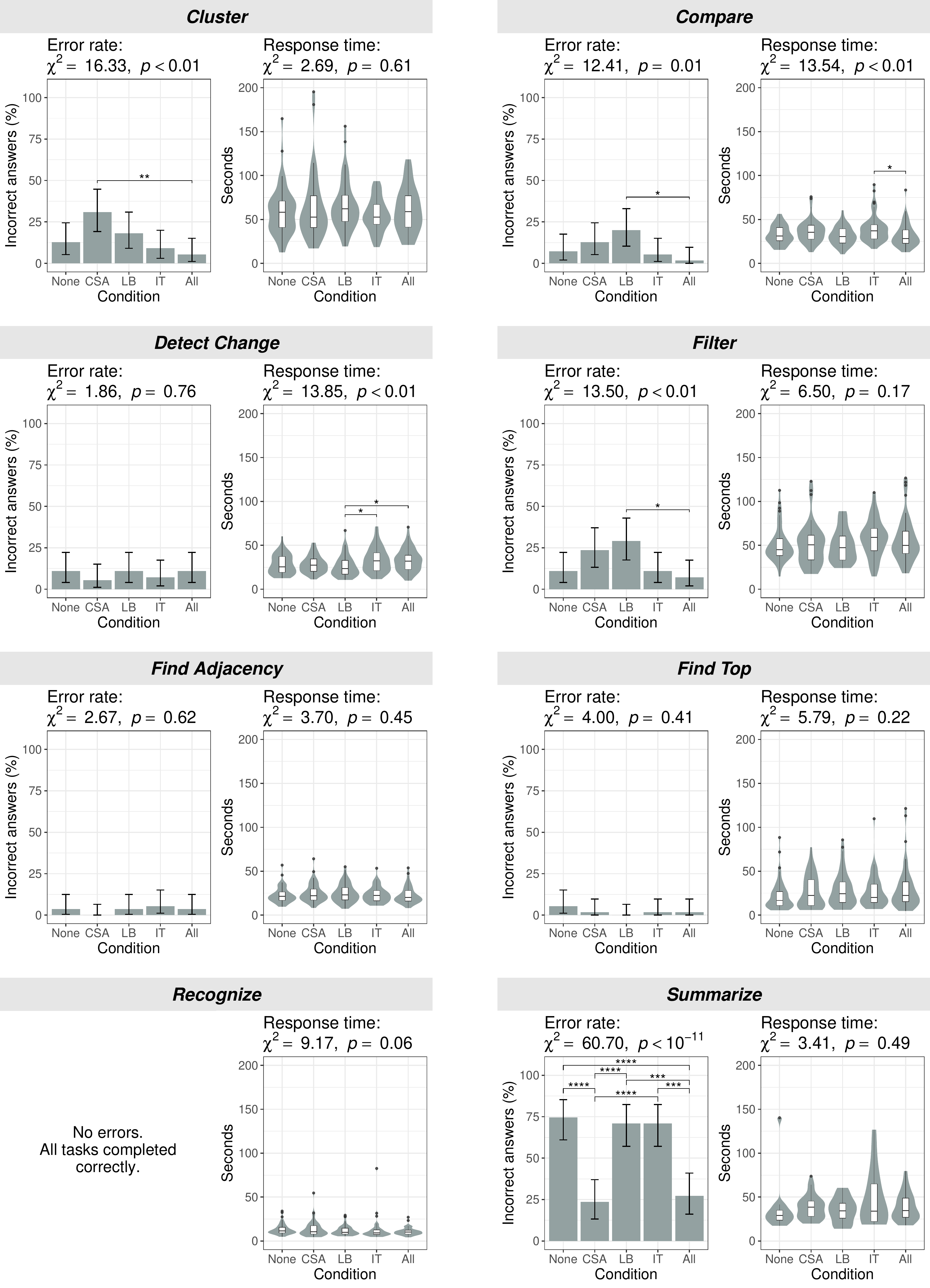}
\caption{Error rates and response time distributions for the cartogram task types in Table~\ref{tab:tasktypes}. We use the following abbreviations for the axis labels. CSA: cartogram-switching animation is the only available interactive feature. LB: linked brushing only. IT: infotip only. Brackets inside the panels indicate significant differences between pairs of conditions at a significance level of 0.05. Asterisks above the brackets indicate $p$-values. $\ast$:~$p$-value $\leq 0.05$, $\ast\ast$:~$\leq 0.01$, $\ast\ast\ast$:~$\leq 0.001$, $\ast\ast\ast\,\ast$:~$\leq 0.0001$. Error bars represent 95\% CIs.}
\label{fig:error_and_time}
\end{figure*}

\subsection{Error Rates and Response Times}

Participants made few errors overall.
The distribution of the number of errors by participant is roughly symmetric and peaked around the mean ($5.5$ errors in $40$ trials) with a range from $0$ to $10$ errors and a standard deviation of $2.3$.
(See Section 2 in the online supplemental text for more summary statistics.)
Because of the narrow distribution, we can regard the group of participants as sufficiently homogeneous to include the responses of all participants for further statistical analysis.

In Fig.~\ref{fig:error_and_time}, we summarize the results of the data analysis outlined in Section~\ref{sec:data_analysis}.
Tabular summaries of the error rates and average response times are in Sections 2 and 3 of the online supplemental text.
In the following list, we provide details about the results for each task type.

\begin{itemize}
\item \textbf{Cluster:}
Judging by a mean error rate of 15.3\%, this task type was of intermediate difficulty.
Interactive features caused a significant effect on the accuracy [$\chi^{2}(4)=16.33$, $p<0.01$], but did not have a significant effect on response times [$\chi^2(4)=2.69$, $p=0.61$].
When participants had access to all interactive features, they
made significantly fewer errors than when they were only allowed to use cartogram-switching animations ($5.5\%$ vs.\ $30.9\%$, odds ratio $0.13$, 95\% CI $[0.02, 0.78]$,
$p=0.01$).
%We did not observe any significant difference between the error rates when either only infotips or only cartogram-switching animations were available ($p=0.07$).

\item \textbf{Compare:}
The participants found this task type slightly easier than \textit{Cluster} (mean error rate 9.5\%).
We observed a significant effect of the interactive features on error rates [$\chi^{2}(4)=12.41$, $p=0.01$] and response times [$\chi^{2}(4)=13.54$, $p<0.01$].
For the error rates, the post-hoc analysis reveals a pairwise difference between the all-features and linked-brushing-only conditions ($1.8\%$ vs.\ $20.0\%$, odds ratio $0.09$, 95\% CI $[0.01, 0.96]$, $p=0.04$).
For the response times, we find a significant difference between the all-features (median $28.1\,\textrm{s}$) and infotip-only conditions (median $36.9\,\textrm{s}$, $\Delta = 7.3\,\textrm{s}$, 95\% CI $[0.9\,\textrm{s}, 14.0\,\textrm{s}]$, $p=0.02$). %but no difference between linked-brushing-only and infotip-only ($p=0.08$).

\item \textbf{Detect Change:}
For tasks of this type, the participants achieved a low mean error rate of 9.1\%, similar to their performance for \textit{Compare}.
Differences between the interactive features did not seem to impact the accuracy; the $p$-value for the main effect is $0.76$ [$\chi^{2}(4)=1.86$].
However, we find a significant effect on response times [$\chi^{2}(4)= 13.85, p<0.01$].
Specifically, participants were significantly faster (median $23.7\,\textrm{s}$) under the linked-brushing-only condition than the all-features condition (median $32.1\,\textrm{s}$, $\Delta = 7.0\,\textrm{s}$, 95\% CI $[0.4\,\textrm{s}, 14.6\,\textrm{s}]$, $p=0.03$) or infotip-only condition (median $32.2\,\textrm{s}$, $\Delta = 6.8\,\textrm{s}$, 95\% CI $[0.4\,\textrm{s}, 14.2\,\textrm{s}]$, $p=0.03$).

\item \textbf{Filter:}
The mean error rate of \textit{Filter} ($16.4\%$) is close to that of \textit{Cluster}.
\textit{Filter} is also similar to \textit{Cluster} in having an effect on accuracy [$\chi^{2}(4)=13.50$, $p<0.01$], but not on response times [$\chi^{2}(4)= 6.50$, $p=0.16$].
The participants' error rates were significantly lower under the all-features condition ($7.3\%$) than under the linked-brushing-only condition ($29.1\%$, odds ratio $0.20$, 95\% CI $[0.04, 0.99]$, $p=0.05$).

\item \textbf{Find Adjacency:}
The participants found this task type easy (mean error rate $3.3\%$). The error rates are low for all conditions, ranging from $0\%$ for linked-brushing-only to $5.5\%$ if there is no interactivity.
The response times also hardly deviate from the median ($21.5\,\textrm{s}$).
Therefore, the interactive features had no significant effect on error rates [$\chi^2(4)=2.67$, $p=0.62$] or on response times [$\chi^{2}(4)= 3.70, p=0.45$].

\item \textbf{Find Top:}
The mean error rate for this task type is similarly low ($2.2\%$) as that for \textit{Find Adjacency} with little variability across different conditions ($0\%$ to $5.5\%$).
We detect no significant effect on accuracy [$\chi^{2}(4)= 4.00, p=0.41$] or on response times [$\chi^{2}(4)= 5.79, p=0.22$].

\item \textbf{Recognize:}
All participants answered every \textit{Recognize} task correctly.
Therefore, no further analysis is needed for the error rates.
This task type also had the shortest median response time ($10.0\,\textrm{s}$) with no significant difference between the experimental conditions [$\chi^{2}(4)= 9.17, p=0.06$].

\item \textbf{Summarize:}
This task type was clearly the most challenging in terms of the mean error rate ($53.5\%$).
Participants' performance depended strongly on the experimental condition.
The error rates range from $23.6\%$ for the cartogram-switching-animation-only condition to $74.5\%$ when there was no interactivity.
The $p$-value for a main effect is accordingly small [$\chi^{2}(4)= 60.70$, $p<10^{-11}$].
The post-hoc analysis reveals that cartogram-switching animations---either as the only interactive feature or in combination with all other features---significantly improved performance compared to all other experimental conditions.
For all pairwise McNemar tests that involve one condition with animations and another without animations, we find $p$-values below $0.01$ and CIs that clearly exclude the null hypothesis of an odds ratio equal to 1.
For example, the 95\% CI for the comparison between cartogram-switching-animation-only and the no-interactivity treatment is $[0.00, 0.35]$. 
In contrast to the clear benefit of animations in terms of accuracy, we do not find a main effect of interactivity on response times [median$=34.1$\,s, $\chi^{2}(4)= 3.41, p=0.49$].
\end{itemize}

\subsection{Performance for Summarize}
\label{subsec:performance_for_summarize}

Because \textit{Summarize} is the task type with the highest error rate, we conducted further data analysis.
We first investigated whether the statistical conclusions depend on the numerical threshold for accepting ``Approximately no change'' as a correct answer.
Then we examined whether different user groups benefit from animations to the same degree.

As we explain in Section~\ref{sec:design}, a pilot study revealed that the minimal detectable area change is $\theta \approx 1\%$.
Repeating the analysis for thresholds equal to $\theta\textsubscript{low} = 0.5\%$ and $\theta\textsubscript{high} = 2\%$, we find that the mean error rate decreases as the threshold increases (from $68.4\%$ for $\theta\textsubscript{low}$ to $34.5\%$ for $\theta\textsubscript{high}$).
This tendency is expected, but it is noteworthy that, regardless of which threshold we choose, the Cochran Q test always rejects the null hypothesis that interactive features had no effect on the error rates.
The numeric results can be found in Section 2 of the online supplemental text.
The post-hoc McNemar tests also always identify the same significant pairwise differences between interactive features.
Hence, our conclusion that animations are a significant help when answering \textit{Summarize} tasks is independent of the exact definition for the minimal detectable area change $\theta$.

We also find evidence that this conclusion is valid for user groups with different levels of prior experience.
We divided participants into two categories based on whether they considered themselves to be familiar with interactive computer graphics ($\geq 4$ on a 5-point Likert scale) during the preliminary questions.
The group with greater familiarity contained 25.5\% of the participants.
We observed that interactivity greatly reduces the error rate in this group from 71.4\% (no features) to 14.3\% (all features).
The error rates for the second group are higher for both conditions, but we find again a clear improvement from 75.6\% (no features) to $31.7\%$ (all features).

The same trend occurs when we divide the participants into two categories based on their general affinity with maps.
To infer their attitude towards maps, we posed the following preliminary question: ``When you encounter the names of unfamiliar locations (e.g., countries, islands, lakes), how frequently do you immediately look them up on a map to find out where they are?''
We dichotomize the participants depending on whether their answer was $\geq 4$ on a 5-point Likert scale.
The group who declared a tendency towards reading maps (34.5\% of the participants) benefited tremendously from the interactive features, decreasing their error rate from 68.4\% (no features) to 26.3\% (all features).
The other group had slightly higher error rates, but we still observe a marked decrease associated with the interactive features (77.8\% without any features, 27.8\% with all features).

Because linked brushing or infotips in isolation do not result in reduced error rates for \textit{Summarize} tasks (Fig.~\ref{fig:error_and_time}), animations seem to be the main reason for the improved performance in the all-features condition.
In Section 2 of the online supplemental text, we show that the positive effect of animations---either as the only available feature or in combination with the other two features---is independent of the participants' confidence using interactive computer graphics and independent of their personal inclination towards reading maps.
Conversely, \textit{Summarize} is the only task type in which animations were associated with significantly improved performance (see Section 2 of the online supplemental text), but error rates for elementary tasks were generally so low that there would not have been much room for improvement.

\subsection{Hypotheses}
In terms of our hypotheses in Section~\ref{subsec:experiment_hypotheses}, \textbf{H1 is partially supported} because we found that cartogram-switching animations improved accuracy in \emph{Summarize} tasks but not in \textit{Detect Change} tasks.
For linked brushing, we observed neither any significant decrease in the error rate nor in the response time compared to the no-interactivity baseline.
Thus, \textbf{H2-a and H2-b are rejected}.
The results for infotips were also inconclusive, so \textbf{H3-a and H3-b are also rejected}.
For seven out of eight task types, the all-features condition showed no significant improvement in accuracy.
The noteworthy exception is \textit{Summarize}.
Hence, \textbf{H4-a is partially supported}, but the all-features performance in \textit{Summarize} was only on par with the cartogram-switching animation, suggesting that the gain in accuracy is caused by that specific feature.
We did not find any significant increase in response times under the all-features condition, so \textbf{H4-b is rejected}.

\begin{figure}
\centering
\includegraphics[width=0.5\textwidth]{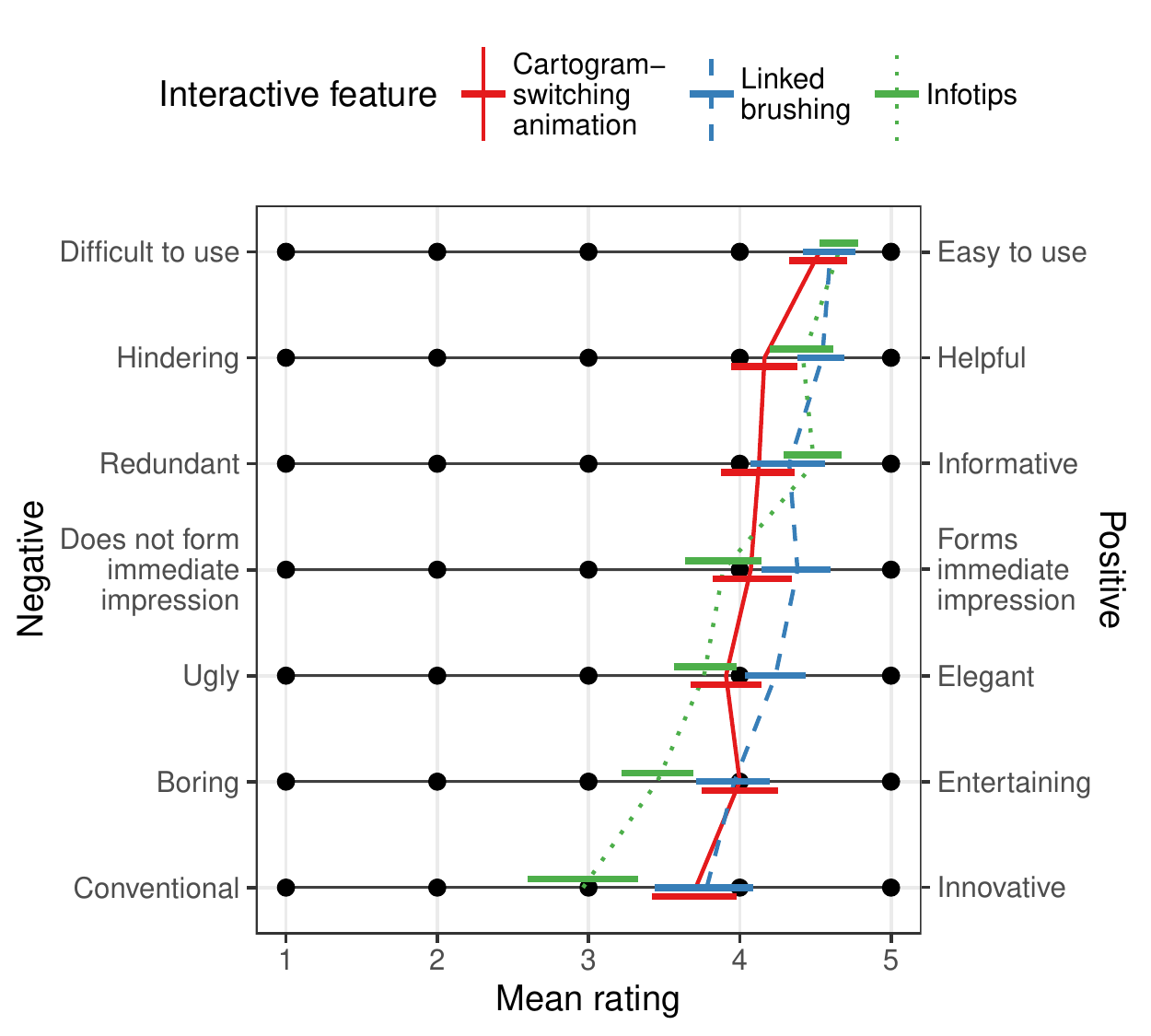}
\caption{Mean ratings in the attitude study conducted at the end of our experiment. Horizontal bars are bootstrap estimates of the 95\% CIs. We sorted the phrase pairs along the vertical axis so that the pair with the overall strongest positive score (``Easy to use'') is at the top.}
\label{fig:attitude_line}
\end{figure}

\subsection{User Preferences}
\label{sec:user_preferences}

In the final part of the experiment, we presented seven pairs of phrases to the participants.
Each pair consisted of two phrases with opposite meaning:
\begin{itemize}
\item Difficult to use -- Easy to use,
\item Does not form immediate impression -- Forms immediate impression,
\item Conventional -- Innovative,
\item Redundant -- Informative,
\item Hindering -- Helpful,
\item Boring -- Entertaining,
\item Ugly -- Elegant.
\end{itemize}

We asked the participants to rate each of the three interactive features (i.e., cartogram-switching animation, linked brushing, infotips) in terms of these phrases on a 5-point Likert scale.
We show the mean rating for each combination of an interactive feature and a phrase pair in Fig.~\ref{fig:attitude_line}.
More information about the distribution can be found in Sections 4 and 5 of the online supplemental text.

The participants gave positive ratings (i.e., mean $>3$) for 20 out of 21 combinations of features and phrase pairs.
Averaged over all interactive features, the positive phrase with the highest rating was ``Easy to use'' (mean rating 4.59).
The positive phrase with the weakest agreement was ``Innovative'' (mean rating 3.48).
Taking the average of all pairs of phrases, linked brushing achieved the highest approval (mean 4.26), closely followed by cartogram-switching animations (4.07) and infotips (3.95).
In summary, participants gave strong subjective feedback about all three features even though only cartogram-switching animations led to objective improvements in performance.

\section{Discussion}

For most tasks in our experiment, we find that the average error rates are below 10\%.
Our results are in line with previous observations that most readers can easily extract information from cartograms even without interactivity~\cite{dent_communication_1975, nunez_hungarian_2015}.
The most notable exception is the task type \textit{Summarize} with an average error rate of 53.5\%.
This task was challenging because participants had to distinguish subtle area differences.
In Fig.~\ref{fig:summarize_example}, for example, the purple zone in the northwest increases by only $1.4\%$ from the 1985 cartogram to the 2015 cartogram.
\textit{Summarize} tasks, therefore, asked participants to assess differences between cartograms more carefully than any other task, resulting in lower overall accuracy.
%We counted the answer ``Approximately no change'' only as correct if the areas differed by no more than 1\%.
Our observation is consistent with results obtained by Kasper \textit{et al.}~\cite{kaspar_empirical_2011}, who also noted that the error rates in their experiment depended strongly on the complexity of the cartogram task.

\subsection{Effect of Cartogram-switching Animations on Performance}
We found that a cartogram-switching animation was an effective way to improve the accuracy in \textit{Summarize} tasks, dramatically reducing the error rate from above 70\% to around 30\%.
Animations may make area changes much easier to detect because the viewer does not have to shift the gaze between two spatially separated cartograms.
Moreover, the smooth transition of the regions' boundaries helps to detect the direction of movement so that it becomes clearer whether the enclosed area expands or contracts.

In contrast to the synoptic \textit{Summarize} tasks, we found that for elementary task types, cartogram-switching animations were less effective, but were not associated with a practically significant decrease in performance.
The error rate of \textit{Cluster} tasks increased nominally if animations were the only available feature (Fig.~\ref{fig:error_and_time}).
However, when we checked the screen recordings, we found that only 17 out of 54 participants actually used cartogram-switching animations when performing a \textit{Cluster} task under the animation-only condition. (For one participant, we have no screen recording of this task.)
The error rate for these 17 participants was almost the same as for the remaining participants (29.4\% with and 29.7\% without using animations).
A two-sample proportion test yields a $p$-value close to 1 with a 95\% CI of $[-26.2\%, 26.8\%]$ for the difference in proportion.
Moreover, the animations were also available under the all-features condition, where the accuracy was higher than in the case of no interactivity.
Therefore, there is no evidence that cartogram-switching worsened the performance for any task type.
In Section 7 of the online supplemental text, we give more details about the frequency with which participants chose to use animations for each task type.

Participants rated cartogram-switching animations as the most entertaining feature (4.00 out of 5).
Especially during \textit{Summarize} tasks, the animations piqued the participants' interest.
In all \textit{Summarize} trials in which an animation was available (i.e., under the cartogram-switching-animation-only and all-features conditions), the screen recordings show that all participants played the animation multiple times.
In many cases, the participants may have repeated the animation to confirm that their answer was correct.
Because the buttons for switching between different cartograms were placed next to each other (e.g., the box labeled ``4'' in Fig.~\ref{fig:cartview} shows the neighboring buttons for GDP and land area), it only needed small hand movements and little motor control to play the animations multiple times.
In addition, participants may have felt encouraged to replay the animation because we told them during the introduction that they could take as long as they needed to answer questions.
Although each animation only lasted for one second, repeated use of this feature is the most likely reason we do not observe a significant speedup for \textit{Summarize} tasks.
Ware~\cite{ware_using_1998} reached a similar conclusion in her cartogram experiment.
We agree that the higher accuracy and the positive subjective ratings for cartogram-switching animations more than outweigh the (statistically insignificant, in our experiment) increase in response times.
Another strong argument in favor of cartogram-switching animations is that they improved the accuracy even for participants who were less confident users of computer graphics or less inclined to read maps.

\subsection{Effect of Infotips on Performance}
For all eight task types, the presence of infotips did not negatively impact the accuracy of the participants' responses in a statistically significant way.
Although in some cases the error rate nominally increased when infotips were the only available feature, none of the pairwise post-hoc tests revealed a significant deterioration compared to the case of no interactivity. In the combination of all three features, infotips did not appear to be detrimental to accuracy either.

Infotips did not significantly affect participants' response times compared to the no-interactivity condition (Fig.~\ref{fig:error_and_time}).
The only effect we observe is that the infotip-only condition was significantly slower than the all-features condition for \textit{Compare} tasks.
Judging from the screen recordings, it appears likely that most participants with access to an infotip performed the \textit{Compare} tasks by reading the numbers in the pop-ups for the two regions mentioned in the question rather than by visually comparing cartogram areas.
Retrieving information from text tends to be slower than from diagrams~\cite{prasad_text_2012}.
Furthermore, after reading the numbers in one pop-up, participants had to shift the mouse to a different region to view the numbers in another pop-up.
Consequently, this strategy was slower than estimating the regions' areas by eye from the cartograms.
An infotip might also have distracted from the task by obstructing parts of the cartograms, thus explaining the observed slowdown.

Consistent with this hypothesis, the participants rated infotips slightly lower in terms of ``forming an immediate impression'' compared to the other interactive features (Fig.~\ref{fig:attitude_line}).
Infotips also received only intermediate ratings halfway between ``conventional'' and ``innovative'' (2.96 on a scale from 1 to 5), presumably because similar mouse-over effects are currently common on many websites.
It is also conceivable that infotips were less popular because the participants had to place the mouse pointer directly on top of the region of interest, which might have felt tedious, especially if the task required activating infotips related to distant regions.
Nevertheless, the overall ratings for infotips are positive, and they do not lead to a significant loss of accuracy compared to the no-interactivity condition for any of the task types (Fig.~\ref{fig:error_and_time}).
We therefore still recommend including an infotip with interactive cartograms.
In our experiment, the pop-up immediately displayed the regions' statistics as soon as the mouse hovered over the map.
The pop-up disappeared only when participants moved the mouse off the map.
A small change in the design of the infotip may make it less obtrusive: if participants can display and hide the pop-up with a mouse click, the infotip will not permanently obstruct space on the map.

\subsection{Effect of Linked Brushing on Performance}
Like infotips, linked brushing did not appear to be detrimental to accuracy. Furthermore, the ratings for linked brushing were positive across the board.
Linked brushing is a subtle, unobtrusive feature that did not cause a significant increase in response time for any task type.
Unlike infotips, linked brushing needs less hand motor control, because the highlighting becomes visible when the mouse pointer moves across the region of interest, but does not need to be placed precisely on top of it.
Our hypotheses H2-a and H2-b, that linked brushing would improve performance in \textit{Find Adjacency}, \textit{Find Top}, and \textit{Recognize} tasks, were not supported.
However, the error rates and response times for these task types were generally low in our experiment, so there was not much room for improvement.
For maps with a larger number of regions than those we used in our experiment, linked brushing may have greater benefits.

\subsection{Effect of Map Presentation, Map Complexity, and Familiarity on Performance}
On all conventional maps, we labeled every administrative unit with a two-letter or three-letter identifier (see Fig.~\ref{fig:cartview}). Space permitting, these labels were placed near the center of the region.
Otherwise, labels were placed outside the map and connected to the corresponding regions by lead-out lines.
Screen recordings indicate that a few participants may have been confused by the density of lead-out lines on some maps, so they may have misidentified some map regions. However, the videos suggest that most participants interpreted region labels and lead-out lines correctly. We conclude that the presentation of these map elements is unlikely to have had a measurable effect on participants' overall performance.

In Section 6 of the online supplemental text, we present statistics about the participants' performance for cartograms of different countries.
In general, the accuracy of the participants' responses did not seem to depend on the country shown on the map.
However, we found clear statistical evidence that the country shown on the map had an influence on the response time.
The responses for Brazil (median $33.9\,\textrm{s}$) as well as for mainland China and Taiwan ($33.2\,\textrm{s}$) were significantly slower than for the United States (median $23.2\,\textrm{s}$).
A regression analysis in Section 6 of the online supplemental text does not find evidence that the median response time increased with the number of administrative units, which was highest on the US map.
Instead, we hypothesize that the response time was mainly influenced by prior familiarity with the conventional map: the US state map is a recognizable icon worldwide, but few Singaporean readers are regularly exposed to maps of Brazilian states.
In this context, it is important to note that our experimental design controlled for country-dependent differences because the displayed countries were a blocking factor (see Section~\ref{sec:design}): for each task type, there were equally many participants assigned to using each combination of a country with an interactive-feature condition.

Although our experiment gave no indication that performance depended on the number of administrative units, we believe that the interactive features work best for the intermediate range we used in our experiment (from 16 for Germany to 49 for the US).
If the units on the screen become too numerous and hence too small, it will become difficult to position the mouse pointer with enough precision to activate the correct infotip or highlight the target region with linked brushing.
We hypothesize that the effectiveness of animations is less dependent on the number of administrative units because animations can always be triggered with the same degree of motor control.
However, a larger number of polygons on the map may reduce an animation's effectiveness owing to the increased cognitive load.
It will be an interesting task for future research to investigate this hypothesis.

\subsection{Generalizability of the Results to Other Cartogram Types}
The cartograms in our experiment were the result of a continuous map projection~\cite{gastner_fast_2018}.
We believe that most of our results apply to other contiguous cartogram types too (e.g., rectilinear~\cite{de_berg_optimal_2010} or mosaic cartograms~\cite{cano_mosaic_2015}) although the absence of a map projection may make it more difficult for viewers to mentally establish the underlying transformation.
If the cartograms are noncontiguous (i.e., regions are displayed by disconnected polygons), the error rates of \textit{Find Adjacency} tasks will presumably be higher than those we found in our experiment.
A switching animation that morphs a noncontiguous cartogram into a conventional map would reveal information about adjacency that is not obvious from a still image.
Therefore, animations may have a stronger positive effect when working with noncontiguous instead of contiguous cartograms.

\subsection{Generalizability to Bivariate Cartograms}

The cartograms in our experiment were univariate maps: we only represented one variable per cartogram (e.g., only population size or only GDP), and areas were the only visual variable that conveyed information about these numbers.
On univariate cartograms, colors can be freely chosen to support readability.
Here we selected fill colors from the 6-class ColorBrewer palette ``Dark2''~\cite{brewer_colorbrewer_2002}, which is designed to work well on liquid-crystal display screens such as those used in our experiment.
We deliberately chose a dark palette so that we could reserve bright colors for the linked brushing effect.
The combination of colors was unlikely to be misconstrued as a way to represent data.
We chose different colors for all neighboring regions, but corresponding regions on the juxtaposed conventional maps and cartograms had the same color so that matching pairs were easy to spot even without linked brushing.

For bivariate cartograms, linked brushing may be more essential for readability.
If colors represent categorical or quantitative data, it may be inevitable that neighboring regions are filled with the same color.
For example, on the classic US presidential election cartogram~\cite{gastner_maps_2005}, one variable (number of electors) is represented by area and the other variable (party affiliation of the electors) by a binary color scheme: red for Republican, blue for Democratic.
On such cartograms, large contiguous swathes of states typically appear in the same color (e.g., the entire South is usually red, the Northeast blue). 
We hypothesize that even elementary cartogram reading tasks become substantially more challenging if the reader cannot take for granted that neighboring regions have distinct colors.
Linked brushing may reduce the challenge posed by tasks of the types \textit{Detect Change} and \textit{Recognize}, which can be answered by finding matching pairs of regions on two juxtaposed maps.
However, linked brushing must be implemented judiciously.
If colors represent data values, we advise avoiding confusion and refraining from changing the color to highlight the region under the mouse pointer.
For bivariate maps, increasing the border thickness promises to be a more effective form of linked brushing than changing the fill color.

Fewer changes are needed to implement the other two interactive features in our study (i.e., infotips and animations) for bivariate cartograms.
As suggested by Nusrat \textit{et al.}~\cite{nusrat_cartogram_2018}, the text in the infotip can simply reveal the data for both variables simultaneously. 
If colors are used to represent one of the thematic mapping variables, a cartogram-switching animation would have to depict area and color changes simultaneously.
Judging from previous experiments~\cite{fish-change-2011}, there is no simple rule whether colors should change abruptly, or whether one should apply tweening to achieve a smooth color transition.
In some cases, animations may in fact make bivariate cartograms entirely superfluous.
As an alternative to representing two mapping variables with two different  visual variables (area and, for example, color), we can make two univariate cartograms and allow users to discover the differences between the cartograms with an animation.\footnote{If both mapping variables are measured in the same units and add up to a meaningful total, Nusrat et al.'s ~\cite{nusrat_cartogram_2018} bivariate pie-chart cartograms make it possible to use color as a single visual variable for both mapping variables. However, additive bivariate data is an exception rather than the rule.}
Animated univariate cartograms allow viewers to concentrate on a single visual variable (area).
We hypothesize that dynamic changes in a single visual variable are easier to detect synoptically than associations between two different visual variables, especially if the animation between two univariate cartograms can be played repeatedly in both directions.
The present study does not allow us to compare animations to bivariate cartograms, but it encourages future research in this direction.

\subsection{Guidelines for Adding Interactivity to Cartograms}

Our experimental results suggest that the need for interactivity depends on the complexity of the task that the cartogram designer has in mind (i.e., whether the task is ``elementary'' or ``synoptic'' according to the typology by Andrienko and Andrienko~\cite{andrienko_exploratory_2006}).

\begin{itemize}
\item
\textbf{Elementary tasks:}
For all seven elementary task types in our experiment (\textit{Cluster}, \textit{Compare}, \textit{Detect Change}, \textit{Filter}, \textit{Find Adjacency}, \textit{Find Top}, and \textit{Recognize}), error rates without any interactivity were very low---below $13\%$. Response times for elementary tasks did not depend significantly on the availability of interactive features.
We conclude that most readers can effectively decode the essential information presented by contiguous cartograms even if the display is static. However, we recommend following Dent's advice~\cite{dent_communication_1975} to always show a cartogram together with a conventional map, which replicates our experimental setting.
\item
\textbf{Synoptic tasks:}
For the only synoptic task in our experiment (\textit{Summarize}), animations clearly had a positive effect.
The error rates were more than halved when an animation was available.
There was no significant effect on the response time, which is partly explained by our observation that animations were always played repeatedly.
If interactive cartograms become less of a rarity in the future, we may find that more experienced users play the animation less often, thus improving efficiency in the long run.
In addition to answering questions about the shown cartograms more accurately, participants also expressed a subjective preference for animations, so we recommend including them regularly if the intended goal is to perform a synoptic task.
The other two interactive features (linked brushing and infotips) neither improved nor worsened the performance.
They can be added as an option but are not strictly necessary.
\end{itemize}

Because we applied the task taxonomy of Nusrat \textit{et al.}~\cite{nusrat_task_2015}, our experiment mainly focused on elementary tasks.
However, one can argue that the primary purpose of cartograms is to perform synoptic tasks.
After all, cartograms of any kind make geometric compromises, either in the shapes or in the contiguity.
If all we want to accomplish is an elementary task, then other map types have the advantage of familiarity to most users.
It would be worth expanding the existing cartogram task taxonomy in the future with a greater variety of synoptic tasks, where cartograms can play to their strengths (e.g., detecting correlations between different variables).
Still, even with only one synoptic task type included in our experiment, we have already found clear evidence in favor of cartogram-switching animations and recommend this feature as a minimum of interactivity.

We have developed a web application (\url{https://go-cart.io/}) that demonstrates our implementations of the interactive features described in this paper~\cite{tingsheng_go-cart.io:_2019}.
Apart from our minimal recommendation of including animations, we combine them on go-cart.io with linked brushing and infotips.
These additional features did not lead to measurable improvements in our experiment, but they were not harmful either.
Because of the generally favorable subjective evaluations by participants in our experiments, our overall recommendation is the full suite of features as implemented by go-cart.io.

\section{Conclusion}

Despite their inherent distortion, cartograms can be an effective tool in the cartographer's toolbox for displaying geospatial statistics.
Previous research has already recommended best practices for displaying printed cartograms~\cite{dent_communication_1975,tobler_thirty_2004,tingsheng_motivating_2020}, such as including a conventional map and a legend as a reference for the reader.
Based on the results of our experiment, we add another recommendation: if displayed electronically, cartograms should be presented with interactivity.
We found that readers performed synoptic \textit{Summarize} tasks much more accurately when they had access to cartogram-switching animations.
Participants also expressed strongly positive opinions about the other two interactive features that we tested in this experiment (i.e., linked brushing and infotips).
They characterized them as easy to use, informative, and helpful.
Therefore, we recommend that cartograms that are displayed electronically---such as those on \url{https://go-cart.io}---should include all three features.
%We have built a website (\url{https://go-cart.io}) that allows users to create and view cartograms presented according to our recommendation~\cite{tingsheng_go-cart.io:_2019}.
We hope that this website will simplify reading and drawing interactive cartograms, similar to the way in which technologies such as Observable~\cite{observable_inc_magic_nodate} and Vega-Lite~\cite{satyanarayan_vega-lite_2017} have simplified the creation of other types of interactive graphics.

% or


%\appendices

%\section{}

\ifCLASSOPTIONcompsoc
  \section*{Acknowledgments}
\else
  \section*{Acknowledgment}
\fi

This work was supported by the Singapore Ministry of Education (grant R-607-000-401-114) and a Yale-NUS College start-up grant (R-607-263-043-121).
We are grateful to Barry J.~Kronenfeld for his comments on the manuscript.
We would like to thank Editage (www.editage.com) for English language editing.

\ifCLASSOPTIONcaptionsoff
  \newpage
\fi



%\bibliographystyle{IEEEtran}

\bibliographystyle{IEEEtran}
\bibliography{IEEEabrv,interactive_cartogram}


\begin{IEEEbiography}[{\includegraphics[width=1in,height=1.25in,clip,keepaspectratio]{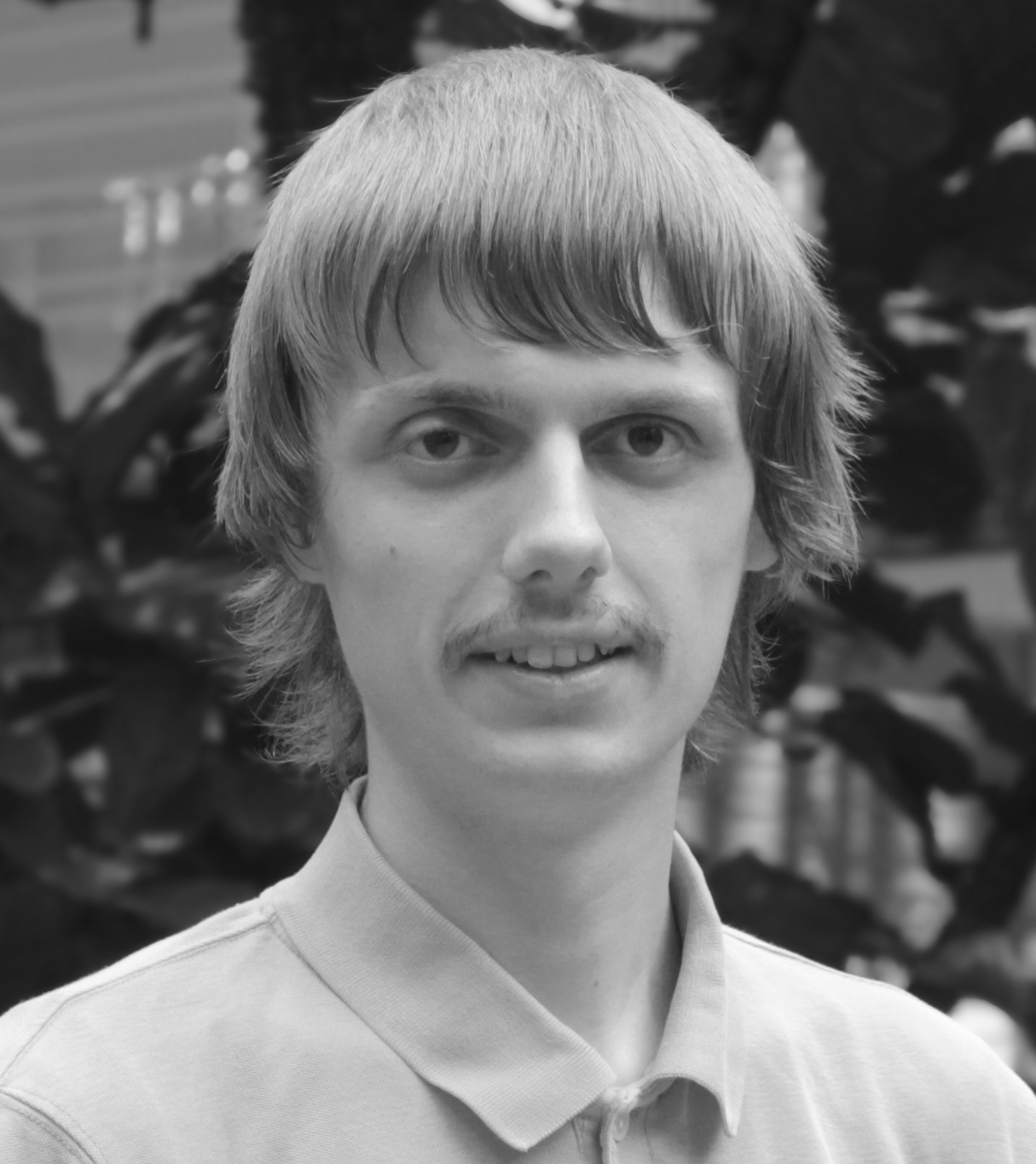}}]{Ian~K.~Duncan}
  is a student at Yale-NUS College in Singapore. He is pursuing a B.Sc.\ in mathematical, computational, and statistical sciences with a minor in philosophy, and will graduate in 2021.
\end{IEEEbiography}

\begin{IEEEbiography}[{\includegraphics[width=1in,height=1.25in,clip]{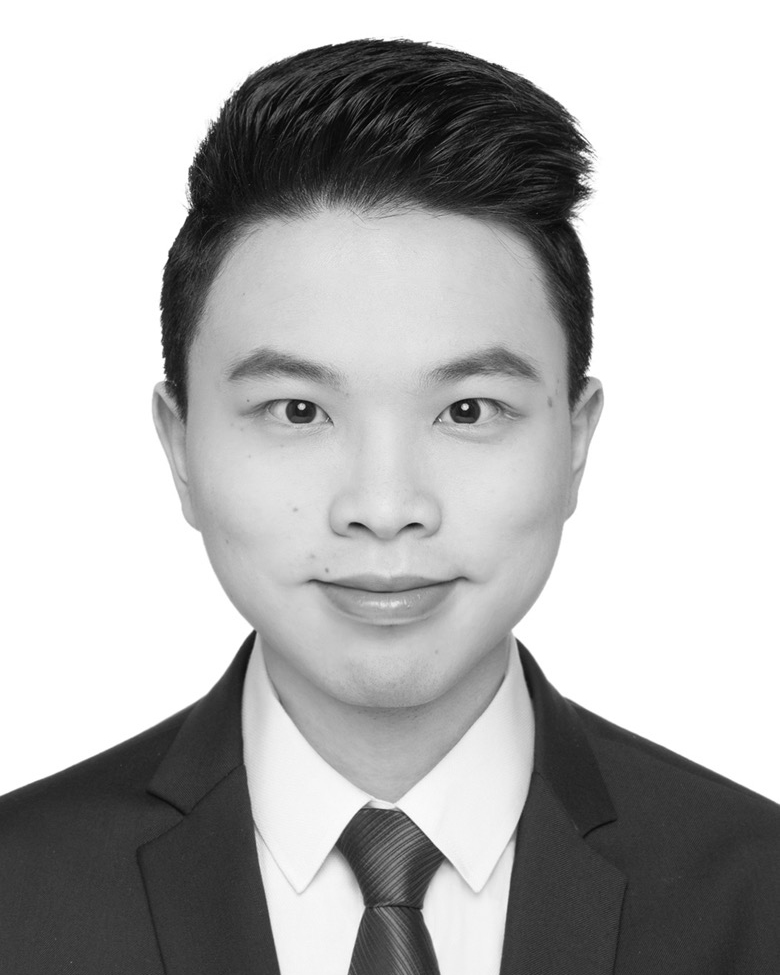}}]{Shi~Tingsheng}
  received his  B.Sc.\ in mathematical, computational, and statistical sciences from Yale-NUS College in 2020. He is currently a software engineer at ByteDance Ltd.
\end{IEEEbiography}

\begin{IEEEbiography}[{\includegraphics[width=1in,clip,keepaspectratio]{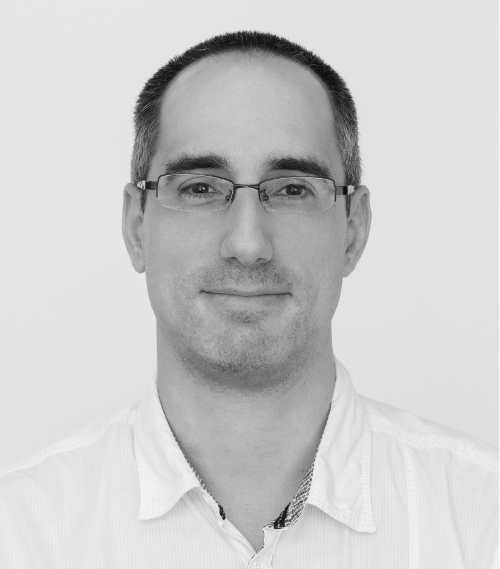}}]{Simon T.~Perrault} obtained his PhD degree in computer science from Telecom ParisTech (France) in 2013.
He is an Assistant Professor in Information Systems Technology and Design at the Singapore University of Technology and Design (SUTD). Previously, he was a Visiting Professor at the Korean Advanced Institute of Science and Technology (Korea), and Assistant Professor at Yale-NUS College (Singapore).
\end{IEEEbiography}

\begin{IEEEbiography}[{\includegraphics[width=1in,height=1.25in,clip,keepaspectratio]{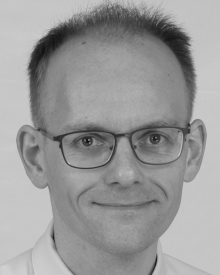}}]{Michael~T.~Gastner}
  received the PhD in physics from the University of Michigan in 2005.
 He held postdoctoral positions at the Santa Fe Institute, the
  University of Oldenburg, and Imperial College London. He 
  is an Assistant Professor for mathematics, computational, and
  statistical sciences at Yale-NUS College in Singapore. Before moving to Singapore, he was a faculty member at the University of Bristol and the Hungarian Academy of Sciences.
\end{IEEEbiography}

%\newpage


%\vfill


\end{document}